\documentclass[]{raa}            
\usepackage{graphicx,times,multirow}
\usepackage{natbib}

\begin{document}

   \title{SWIFT J1749.4$-$2807: A neutron or quark star?}

 \volnopage{ {\bf 2010} Vol.\ {\bf 9} No. {\bf XX}, 000--000}
   \setcounter{page}{1}

   \author{Junwei Yu
   \and Renxin Xu
   }
   \institute{School of Physics and State Key Laboratory of Nuclear Physics and
   Technology,\\
Peking University, Beijing 100871, China.~~~ {\it J.W.Yu@pku.edu.cn,
r.x.xu@pku.edu.cn}\\\vs \no
   {\small Received [year] [month] [day]; accepted [year] [month] [day] }
}

\abstract{We investigate an unique accreting millisecond pulsar with
X-ray eclipses, SWIFT J1749.4$-$2807 (hereafter J1749), and try to
limit the binary system by various methods including that of the
Roche lobe, the mass-radius relations of both a main sequence (MS)
and a white dwarf (WD) companion stars, as well as the measured mass
function of the pulsar. The calculations are based on the assumption
that the radius of the companion star has reached its Roche radius
(or at 90\%), but the pulsar's mass has not been assumed to be a
certain value. Our results are as follows. The companion star should
be a MS. For the case that the radius equals to its Roche one, we
have a companion star with mass $M\simeq 0.51 M_{\odot}$ and radius
$R_{\rm c}\simeq 0.52R_{\odot}$, and the inclination angle is
$i\simeq 76.5^{\circ}$; for the case that the radius reaches 90\% of
its Roche one, we have $M\simeq 0.43M_{\odot}$, $R_{\rm c}\simeq
0.44R_{\odot}$ and $i\simeq 75.7^{\circ}$. We also obtain the mass
of J1749, $M_{\rm p}\simeq 1M_\odot$, and conclude that the pulsar
could be a quark star if the ratio of the critical frequency of
rotation-mode instability to the Keplerian one is higher than $\sim
0.3$. The relatively low pulsar mass (about $\sim M_\odot$) may also
challenge the conventional recycling scenario for the origin and
evolution of millisecond pulsars. The results presented in this
paper are expected to be tested by future observations.
 }

   \authorrunning{Yu \& Xu}            
   \titlerunning{SWIFT J1749: A neutron or quark star?}  
   \maketitle

\section{Introduction}

One of the daunting challenges nowadays is to understand the
fundamental strong interaction between quarks in the low-energy
limit, i.e., the non-perturbative QCD (quantum chromo-dynamics).
This unsolved problem results in the uncertainty to determine the
nature of pulsar-like compact stars: either normal neutron stars or
quark stars (e.g., Lattimer \& Prakash 2004; Xu 2009).
Nevertheless, compact stars may provide as astrophysical
laboratories to understand the non-perturbative QCD in turn, and the
pulsar mass and radius distribution would have important
implications for the nature of pulsar and for the states of matter
at supra-nuclear density.
Certainly it is a very difficult task to determine precisely the
masses of pulsars even in binary systems because of unknowing
inclination angle, unless for general relativistic binaries in which
many post-Keplerian parameters observed are applied.

An amazing system, SWIFT J1749.0-2807 provides us a perfect
opportunity for measuring the mass of pulsar.
J1749 is in a binary system, and the pulsar is in the phase of
accreting matter from and is also eclipsed by its companion star.
It was discovered in June 2, 2006 (Schady et al. 2006).  J1749 is
the first eclipsing accretion-powered millisecond X-ray pulsar
(AMXP) system. Observations set an upper limit distance to this
system of $6.7 \pm 1.3$ kpc (Wijnands et al. 2009). The rotation
period of the pulsar is $\simeq 1.93$ ms and the eclipse by the
companion star centered at the orbital phase of superior conjunction
of the pulsar (Markwardt et al. 2010). The eclipse duration is
$\simeq 2065$ s (corresponding to an eclipse half-angle of $\simeq
11.7^{\circ}$) (Altamirano et al. 2010). Unfortunately, no optical
counterpart has been identified yet, with a $3\sigma$ upper limit in
the I-band of $>$ 19.6 (Yang et al. 2010).

In this {\it Letter}, we are investigating the properties of the
companion star in the case of both main sequence star (MS) and white
dwarf (WD), and trying to obtain the pulsar mass. The calculation
details are presented in \S2 and \S3, and the paper is summarized in
\S4. Through detail calculations, we find that the companion star
could be a main-sequence but cannot be a white dwarf.

\section{To understand the nature of J1749 by observations}

It is known from observations that J1749 is similar to other
low-mass X-ray binary systems driven by disk accretion (Markwardt \&
Strohmayer 2010), and the observational facts of J1749 are
summarized in Table \ref{J1749} taken from Altamirano et al. (2010).
\begin{table}
\caption{The timing parameters of J1749 to be used in our
calculations, taken from Table 1 of Altamirano et al. (2010)}
\label{J1749}
\begin{tabular}{lc}
\hline \hline
Parameter & Value  \\
\hline \\
Orbital Period, $P_{\rm orb}$ (days) & 0.3673696(2) \\
Projected semi major axis, $a_{\rm p} \sin i$ (lt-s) & 1.89953(2) \\
Eccentricity, $e$ (95\% c.l.) & $< 2 \times 10^{-5}$ \\
Spin frequency 1st overtone, $\nu_{0}$ (Hz) & 1035.8400279(1) \\
pulsar mass function, $f_{\rm p}$ ($M_{\odot}$) & 0.0545278(13)\\
\hline
\end{tabular}
\end{table}
We assume that the companion star should reach its Roche lobe which
is approximated by (Eggleton 1983)
\begin{equation}\label{roche lobe 1}
R_{\rm L}=a \times \frac{0.49 \cdot q^{2/3}}{0.69 \cdot q^{2/3} + \ln(1+q^{1/3})} ,
\end{equation}
where $a$ is the semi-major axis of the system and $q=$$M_{\rm
c}/M_{\rm p}$ is the ratio between the companion star and the pulsar
masses.
Importantly, J1749 is the unique system showing eclipse up-to-date.
In such an eclipsing system, from geometrical considerations,
$R_{\rm L}$ is also related to the inclination $i$ and the eclipse
half-angle $\phi \simeq 11.7^{\circ}$ by (Chakrabarty et al. 1993,
Altamirano et al. 2010)
\begin{equation}\label{roche lobe 2}
R_{\rm L}=a \times \sqrt{\cos^{2}i + \sin^{2}i \cdot \sin^{2} \phi},
\end{equation}
where the eccentricity of the system is chose to be zero. With the
above two equations, we can obtain a rough lower limit of this
system according to $\ln (1+q^{1/3}) > 0$, which is about $i>
46^{\circ}$, and we will then consider only large inclination angles
in this paper.

As we all known that the information of the orbital inclination of
binary system is crucial to understanding the properties of its
member stars. Unfortunately, determining the inclination is a
fantastically difficult work. Fortunately, for the case of eclipsing
J1749 system, with the assumption that the radius of the companion
star has reached its Roche lobe, we can investigate the relation
between the mass of pulsar, $M_{\rm p}$ and the radius of the
companion star, $R_{\rm c}$ (namely, $R_{\rm L}$). This relation can
be obtained as follow.  According to Eq. (\ref{roche lobe 1}) and
Eq. (\ref{roche lobe 2}), with different fixed $i$, there will be
different $q$. Additionally, the Kelper's third law is
\begin{equation}\label{kepler3}
 a=[\frac{G (M_{\rm p}+M_{\rm c}) P_{\rm orb}^{2}}{4 \pi^{2}}]^{\frac{1}{3}}=1.47716[x(1+q)]^{\frac{1}{3}} ~\; R_{\odot},
\end{equation}
where $P_{\rm orb}= 8.82$ hr is the orbital period of J1749 system,
$x=$$M_{\rm p}/M_{\odot}$ is the mass of pulsar in unit of solar
mass and $R_{\odot}$ is radius of the sun. With certain $i$ and
equations (\ref{roche lobe 2}) and  (\ref{kepler3}), the relation
between the mass of pulsar and the radius of the companion star is
obtained.

\begin{figure}
\centering
  \includegraphics[scale=0.3]{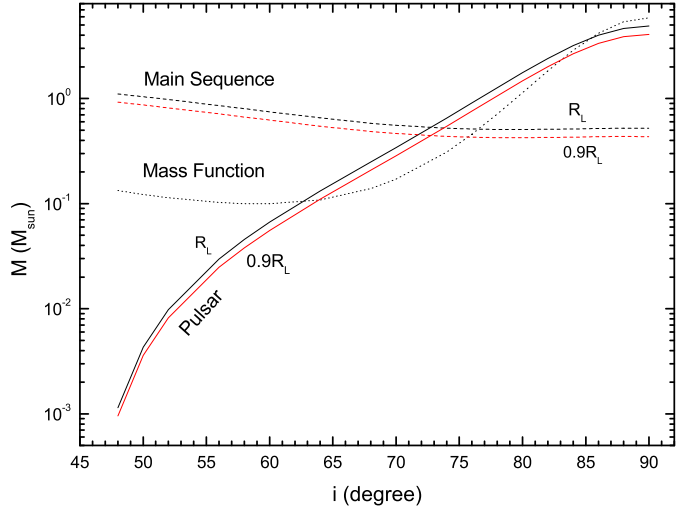}
  \caption{The masses of both pulsar (solid lines) and its companion
  (dashed lined) as function of
  orbital inclination angle. The constraint from the mass function is also
  drawn (dotted lines). Two cases that the companion reaches its Roche lobe
  ($R_{\rm MS}= R_{\rm L}$) or not ($R_{\rm MS}= 0.9 R_{\rm L}$) are considered.
  In this plot, the companion star is assumed to be a MS.}
  \label{MSI}
\end{figure}

Another consideration comes from the mass-radius relation of
companion star, which is empirically given for ZAMSs (zero-age
main-sequences) as (Schmidt-Kaler 1982)
\begin{eqnarray}\label{MRMS}
\log(\frac{R}{R_{\odot}})=0.640 \log(\frac{M}{M_{\odot}}) + 0.011 &(0.12<\log(\frac{M}{M_{\odot}})<1.3), \\
\log(\frac{R}{R_{\odot}})=0.917 \log(\frac{M}{M_{\odot}}) - 0.020 &(-1.0<\log(\frac{M}{M_{\odot}})<0.12).
\end{eqnarray}
The above mass-radius relation for ZAMS is consistent with the
statistical results presented by Demircan \& Kahraman (1991).
What we are interested in is the intersection points of $R_{\rm L}$
and the radius determined by the above mass-radius relation with
fixed inclination angle. It is very meaningful to find out all of
these intersection points with available inclination angles in order
to constrain on properties of both the pulsar and the companion
star. The result is showed in Fig.~\ref{MSI}.
\begin{figure}
\centering
  \includegraphics[scale=0.3]{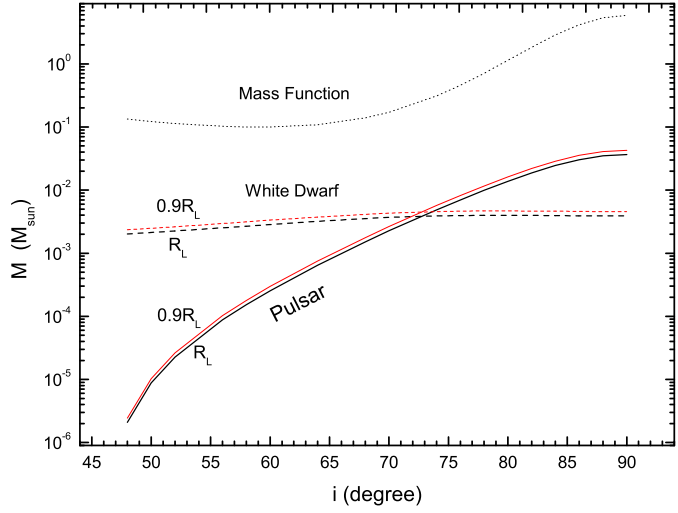}
  \caption{Same as in Fig.~\ref{MSI}, but for white dwarf companion.}
  \label{WDI}
\end{figure}

The third constraint for the companion star comes from the measured
mass function of pulsar. The mass function for binary system is
(Altamirano et al. 2010)
\begin{equation}
f_{\rm p}=\frac{(M_{\rm c} \sin i)^{3}}{(M_{\rm p}+M_{\rm
c})^{2}}=0.0545278M_\odot. \label{MF}
\end{equation}
Because of unknowing about the orbital inclination, the mass
function sets only the minimum companion mass. Nonetheless, if we
cover all of the available inclination angle, the measured mass
function could give another constraint of the companion star.
Combining this constraint and that from previous geometrical
consideration, we may obtain unique mass ratio $q$ and thus the
companion mass, as well as the inclination angle. The resulting
curve is also plotted in Fig.~\ref{MSI} for MSs. We can see in
Fig.~\ref{MSI} only one intersection point exists between the mass
function curve and the Roche curve, which means that the companion
star could be a MS with mass of $M_{\rm c} \simeq 0.512 M_{\odot}$.
Consequently, the mass of the pulsar is $M_{\rm p} \simeq 0.989
M_{\odot}$ and the orbital inclination angle is $i \simeq
76.5^{\circ}$.

The above calculation is based on the assumption that the radius of
the companion star equals to its Roche radius. Keep in mind that for
an evolving MS star, its radius will be a little bit larger than
that of a ZAMS at the same age. This means that the actual radius of
the companion star is a little bit larger than the Roche lobe,
$R_{\rm L}$. Because of this, calculation that the radius of the
ZAMS companion star is a little bit smaller than its Roche lobe is
necessary. In our calculation we chose that the radius of the
companion star is about $90\%$ of its Roche one, namely $R_{\rm
c}=0.9 R_{\rm L}$. The corresponding curve is also plotted in
Fig.~\ref{MSI}. In this case, we have $M_{\rm c} \simeq
0.43M_{\odot}$, $i \simeq 75.7^{\circ}$ and $M_{\rm p} \simeq 0.715
M_{\odot}$.

We also did the same calculation for the case of a WD companion,
whose the mass-radius relation is (Shapiro \& Teukolsky 1983)
\begin{equation}\label{WDMR}
M=0.7011 (\frac{R}{10^4~{\rm km}})^{-3} (\frac{\mu_{e}}{2})^{-5} ~\;
M_{\odot},
\end{equation}
where $\mu_{e}=A/Z$ is the mean molecular weight per electron. For
He-WDs and CO-WDs, $\mu_{e}$ is about 2. Rewriting equation
(\ref{WDMR}), we have
\begin{equation}\label{MRWD}
R_{\rm WD}=0.0161846R_\odot~(\frac{M_{\rm WD}}{M_{\odot}})^{-1/3}
=0.0161846 (qx)^{-1/3}~R_{\odot}.
\end{equation}
The result is showed in Fig.~\ref{WDI}. From Fig.~\ref{WDI}, we can
see that there is no intersetion between the mass function curve and
the Roche curve of WD, which means that WD can be ruled out as the
candidate for the companion star.

\section{Constraint on the equation of state of compact stars}
Once obtaining the mass information of the pulsar J1749, one may
investigate the nature of pulsar, namely a neutron or quark star.
Different equation of state (EoS) can result in different
mass-radius relations, and we may calculate the Keplerian
frequencies of J1749 in various EoS models.
In Newtonian gravity, the Keplerian frequency of the pulsar is gave
by
\begin{equation}
\Omega_{\rm K}=\sqrt{\frac{G M_{\rm p}}{R^{3}}}=11549.9
\sqrt{\frac{x}{r^{3}}}  ~\; {\rm s^{-1}}, \label{KF}
\end{equation}
where $G$ is the gravitational constant, $x=M_{\rm p}/M_{\odot}$ and
$r=R/(10^{6}$ cm).
The mass-radius relations resulted from typical EoSs are showed in
Lattimer (2007, Fig.~2 there), and the calculated Keplerian
frequencies in typical neutron star model are listed in Table
\ref{KFC}.
\begin{table}
\caption{The Kelper frequency  $\Omega_{\rm K}$ of the pulsar J1749
under the assumption that both $R_{\rm c}$$=$$R_{\rm L}$
 and $R_{\rm c}$$=$$0.9 R_{\rm L}$ for a MS companion. The angular
 frequency is fixed to be $\Omega$$= 3254.06$ ${\rm s^{-1}}$. Pulsar's radii for different EoSs are
 took from Lattimer (2007).}
 \label{KFC}
  \begin{tabular}{ccccccc}
    \hline\hline
    &\multicolumn{3}{c}{$R_{\rm c}$$=R_{\rm L}$, $ M_{\rm p}=$$0.989 M_{\odot}$}&\multicolumn{3}{c}{$R_{\rm c}$$=0.9 R_{\rm L}$, $ M_{\rm p}=$$0.715 M_{\odot}$} \\
    \cline{2-7}\\
EoS & $R/(10 \: {\rm km})$  & $\Omega_{\rm K}$ $({\rm s^{-1}})$ &$\Omega$$/$$\Omega_{\rm K}$&$R/(10 \: {\rm km})$  & $\Omega_{\rm K}$ $({\rm s^{-1}})$ &$\Omega$$/$$\Omega_{\rm K}$ \\
    \hline
    MS0&1.471&6438.42&0.505413&1.457&5554.85&0.585805 \\
    MS2&1.432&6703.22&0.485447&1.417&5791.71&0.561848\\
    PAL1&1.4   &6934.35&0.469267&1.403&5878.61&0.553542\\
    FSU &1.325&7531.37&0.432067&1.368&6105.66&0.532958\\
    PAL6&1.225&8472.15&0.384089&1.3&6590.93&0.493718\\
    MPA1&1.225&8472.15&0.384089&1.218&7267.59&0.447749\\
    AP3&1.2&8738.28&0.372391&1.189&7535.09&0.431854\\
    AP4&1.139&9449.58&0.34436&1.143&7994.52&0.407037\\
    SQM1&0.986&11732.3&0.277359&0.889&11654.9&0.279201\\
    SQM3&0.829&15218.3&0.213825&0.736&15471.9&0.21032\\
    \hline
  \end{tabular}
\end{table}

According to general relativity, the rotation-mode (i.e., $r$-mode)
instability could be excited in fast relativistic stars and they may
spin down effectively through radiating gravitational wave
(Andersson et al. 2009).
Certainly it depends on detail EoS and micro-physics about viscosity
to determine the $r$-mode instability window.
Nevertheless, one could calculate a critical angular frequency as a
function of temperature, $\Omega_{\rm c}(T)$, in certain star
models.
The low limit of of critical frequency could be $<0.1\Omega_{\rm K}$
(Andersson et al. 2009), and the pulsar J1749 could be a quark star
if the critical frequency is higher than $\sim 0.3\Omega_{\rm K}$
according to Table 2, since quark stars can sustain faster spin than
neutron stars.

\section{Conclusions and Discussions}

According to our calculations, we address that the companion star
should be a MS with $\sim 0.5 M_{\odot}$ and the inclination angle
is about $76^{\circ}$. Assuming a distance of 7 kpc, the peak
outburst luminosity was about $1.8 \times 10^{36} \: {\rm erg/s}$
(Ferrigno et al. 2010). The pulsar accretes matter from its MS
companion star, and we can estimate the timescale of this accreting
system to be order of $10^{9}$ yr.

Under the assumption that the companion star has reached its Roche
lobe, with the observed mass function and the mass-radius relations
of both MS and WD, we investigate all of the possible solutions of
available inclination angle as well as the corresponding companion
star and pulsar masses. All the results are presented in Table
\ref{MR}.
\begin{table}
\caption{The deduced parameters of the J1749 system.} \label{MR}
\begin{tabular}{lcc}
\hline \hline
  &$R_{\rm c}=$$R_{\rm L}$ &$R_{\rm c}=$$0.9 R_{\rm L}$ \\
  \hline
   $i$ & $76.5^{\circ}$&$75.7^{\circ}$\\
   $R_{\rm c}$($R_{\odot}$) & 0.517 &0.440\\
  $M_{\rm c}$($M_{\odot}$) &0.512 &0.430  \\
  $M_{\rm p}$($M_{\odot}$) &0.989 &0.715  \\
  $B$(G) &$\sim 10^{7}$ & $\sim 10^{7}$\\
  \hline
\end{tabular}
\end{table}
The pulsar mass is $\simeq 1.0 M_{\odot}$ which is about the same as
the lowest one determined up-to-date, SMC X-1, which is
$1.06_{-0.10}^{+0.11}M_{\odot}$ (van der Meer et al. 2007).
According to the mass of the MS companion, we predict the luminosity
of the binary system is $\sim 0.09L_\odot$ in optical band, and we
expect to detect the companion by the following up observations.

From Garcia et al. (2001, Fig.~1), the minimum luminosity of J1749
should be the order of $10^{-5}$$L_{\rm Edd}$, namely $10^{33} {\rm
erg\;s^{-1}}$. One possible reason that J1749 becomes undetectable
by {\it Chandra}\footnote{Private communication with D. Altamirano.}
could be that the pulsar is now in a propeller phase, and we could
then estimate the magnetic field of the pulsar.
The pulsar may manifest itself as transient X-ray source if the
magnetospheric radius is approaching the corotation radius. The
magnetospheric radius is
\begin{equation}\label{RM}
r_{\rm m}=(\frac{B^{2} R^{6}}{\dot{M} \sqrt{2 G M_{\rm p}}})^{2/7}
\simeq 1.3 \times 10^{11} \mu_{30}^{4/7} (\dot{M}_{10})^{-2/7}
(\frac{M_{\rm p}}{M_{\odot}})^{-1/7} ~\; {\rm cm},
\end{equation}
where $\mu_{30}=\mu/(10^{30}\;{\rm G\;cm^{3}})$ is the magnetic
momentum ($\mu=B R^{3}/2$) in unit of $10^{30} \: {\rm G\;cm^3}$,
and $\dot{M}_{10}=\dot{M}/(10^{10} \;{\rm g\;s^{-1}})$ is the
accretion rate in unit of $10^{10} \; {\rm g\;s^{-1}}$. The
corotation radius is given by
\begin{equation}\label{RCO}
r_{\rm co}=(\frac{G M_{\rm p}}{4 \pi^{2}})^{1/3} P^{2/3}\simeq
1.5\times 10^8({M_{\rm p}\over M_\odot})^{1/3}P^{2/3}~\;{\rm cm}.
\end{equation}
From $r_{\rm m}\sim r_{\rm co}$, we could estimate the magnetic
field of J1749 to be
\begin{equation}\label{B}
B \sim 1.45 \times 10^{7} (\frac{M_{\rm p}}{M_{\odot}})^{5/6}
\dot{M}_{10}^{1/2} P^{7/6} R_{6}^{-3} \sim 1\times 10^7R_6^{-3}~{\rm
G},
\end{equation}
where $R_{6}=$$R/(10^{6}\;{\rm cm})$. Therefore, J1749 may have
relatively weak surface magnetic field.

An accretion-driven millisecond pulsar with $\sim 1 M_\odot$ mass
would have profound implications to neutron star physics.
Millisecond pulsars are supposed to be more massive than normal
pulsars in the conventional recycling scenario. However, the mass of
J1749 is much lower than the generally accepted pulsar mass $\sim
1.4M_\odot$. A similar case of another accreting and relatively low
mass compact star, EXO 0748-676, was also discussed in Xu (2003).
These facts may also challenge the standard model for the origin and
evolution of millisecond pulsars, besides the arguments from
population synthesis (Lorimer et al. 2007).
Additionally, considering the $r$-mode instability, we also estimate
the ratio of the real angular frequency to the critical angular
frequency, and find that neutron stars could hardly spin so fast
(spin period $\sim 2$ ms) but quark stars can due to
self-confinement and solidification (Xu 2005).

\section*{Acknowledgments}
We are grateful to Dr. C. B. Markwardt and Dr. D. Altamirano for
communicating the observations of the companion star, and also
acknowledge that Prof. Shuangnan Zhang suggests to discuss the
magnetic field of J1749. We thank the members at PKU pulsar group
for helpful discussions.
This work is supported by NSFC (10778611, 10973002), the National
Basic Research Program of China (grant 2009CB824800).

\label{lastpage}

\end{document}